\documentclass[12pt]{article}
\usepackage[english]{babel}
\usepackage{amsmath,amssymb,amsfonts,textcomp}
\usepackage{color}
\usepackage{calc}
\usepackage{hyperref}
\hypersetup{colorlinks=false, linkcolor=blue, filecolor=blue, pagecolor=blue, urlcolor=blue}
\newcommand\liststyleWWviiiNumi{%
\renewcommand\theenumi{\arabic{enumi}}
\renewcommand\theenumii{\arabic{enumii}}
\renewcommand\theenumiii{\arabic{enumiii}}
\renewcommand\theenumiv{\arabic{enumiv}}
\renewcommand\labelenumi{\theenumi.}
\renewcommand\labelenumii{\theenumii.}
\renewcommand\labelenumiii{\theenumiii.}
\renewcommand\labelenumiv{\theenumiv.}
}
\newcommand\liststyleWWviiiNumii{%
\renewcommand\theenumi{\arabic{enumi}}
\renewcommand\theenumii{\arabic{enumii}}
\renewcommand\theenumiii{\arabic{enumiii}}
\renewcommand\theenumiv{\arabic{enumiv}}
\renewcommand\labelenumi{\theenumi.}
\renewcommand\labelenumii{\theenumii.}
\renewcommand\labelenumiii{\theenumiii.}
\renewcommand\labelenumiv{\theenumiv.}
}
\makeatletter
\newcommand\ps@Standard{%
\renewcommand\@oddhead{}%
\renewcommand\@evenhead{\@oddhead}%
\renewcommand\@oddfoot{\hfil\thepage\hfil}%
\renewcommand\@evenfoot{\hfil\thepage\hfil}%
\setlength\paperwidth{8.5in}\setlength\paperheight{11in}\setlength\voffset{-1in}\setlength\hoffset{-1in}\setlength\topmargin{0in}\setlength\headheight{12pt}\setlength\headsep{0.8035in}\setlength\footskip{12pt+0cm}\setlength\textheight{11in-0in-1in-0.8035in-12pt-0cm-12pt}\setlength\oddsidemargin{1in}\setlength\textwidth{8.5in-1in-1in}
\renewcommand\thepage{\arabic{page}}
\setlength{\skip\footins}{0.0398in}\renewcommand\footnoterule{\vspace*{-0.0071in}\noindent\textcolor{black}{\rule{0.25\columnwidth}{0.0071in}}\vspace*{0.0398in}}
}
\makeatother

\title{Are we observing Lorentz violation in gamma ray bursts?}

\author{Theodore G.\ Pavlopoulos \\U.S. Space and Naval Warfare Systems Center, Code 2361\\ San Diego, CA 92152{}-5001, U.S.A. \\ email: \url{pavlopoulos@cox.net}}

\begin{document}

\maketitle

\begin{abstract}
From recent observations of gamma{}-ray bursts (GRBs), it appears that
spectral time lags between higher{}-energy gamma rays photons and
lower{}-energy photons vary with energy difference and time (distance)
traveled. These lags appear to be smaller for the most luminous (close)
bursts but larger for the fainter (farther away) bursts. From this
observation, it has been suggested that it might be possible to
determine the distance ($L$) these bursts have traveled from these time
lags alone, without performing any red{}-shift measurements. These
observed spreads (dispersion) of high{}-energy electromagnetic pulses
of different energies with time contradict the special theory of
relativity (STR). However, extended theories (ET) of the STR have been
developed that contain a dispersive term, predicting the above
observations. An example of such an ET is presented, allowing us to
derive a relationship between time lags of gamma rays of different
energies and distance $L$ traveled from their origin. \ In addition, this
theory predicts the origin of X{}-ray flashes.
\end{abstract}

\section{I. Introduction}

{\selectlanguage{english}

This article provides a short overview on gamma ray bursts (GRBs), including the observation of X-ray flashes (XRFs).
The unusual behavior in the time separation between two gamma ray pulses of different energy from GRBs is explained
 as a Lorentz violation.

  A short introduction to the STR is presented. It is well known that relativistic quantum mechanics is plagued with divergence problems. For example, this theory is not able to calculate the mass-spectrum of elementary particles such as the electron, proton, neutron, etc. This suggests that either the STR or quantum mechanics
are incomplete physical theories. To remove these divergences, an extended theory (ETs) was proposed by Pavlopoulos \cite{1,2}.
This theory is not a new theory that competes with the STR, but extends the STR. 
It contains an additional term in the wave equation that describes the propagation of electromagnetic waves. 
This term contains a universal length $\ell_0$ as a second invariant. 

The resulting superluminal dispersion relation regulates a point source at $r=0$.

Heisenberg \cite{3} and others emphasized that a so-called universal (or minimal) length $\ell_0$ might play an important 
role in physics. By somehow incorporating this length into physics, it was hoped, might eliminate above difficulties.

Until the 1980s, it was speculated that the universal (or minimal)
length $\ell_0$, if it existed, was somehow
 related to the classical radius of an electron $R_e=e^2/m_ec^2=2.82\times 10^{-13}$cm, with  $m_e$ the mass of an electron, or the Compton
 wavelength $\lambda_C=1.32 \times 10^{-13}$ cm, $m_Pc=h/ \lambda_C$ with $m_P$ the proton
mass.}

{\selectlanguage{english}
During the last 30 years, several observations were made in physics that
resisted explanation. From the late 1980s forward, violations of the
Lorentz invariance have been widely studied in theoretical physics.
 For example, Kostelecky and Samuel \cite{4} reported the appearance of a minimal length in string theory. The following years, more dispersion relations were reported in the literature. According to Colladay and Kostelecky \cite{5}, all these
dispersion relations are contained in the effective field framework
called the standard model extension.}

{\selectlanguage{english}
Significantly, these newer studies put the actual size of the universal
length $\ell_0$ in the range of the
Planck{}-length $L_p$, with $L_p \equiv (h G/c^3)^{1/2} = 1.6 \times 10^{-33}$ cm  and $G$ the gravitational
constant. Since $L_p$ is formed from the universal
constants $h$, $G$, and $c$, $L_p$ is also a universal constant.
These newer theories containing this constant are often referred to as
quantum gravity (QG). Because of the smallness of the length scale
$L_p$, it is expected that effects of QG are only
observable at ultrahigh energies (Planck energy $E_P = h/cL_p \approx 10^{19} GeV$). }

{\selectlanguage{english}
It is not the objective of this letter to discuss these developments of
QG in detail. Astronomical (including GRBs) observations, experimental
and theoretical efforts on Lorentz violations have been discussed in some
detail and references provided in the review articles by Magueijo \cite{6},
``New varying speed of light theories,'' the ``Lecture Notes in
Physics, Quantum Gravity Phenomenology,'' by Jacobson, Mattingly, and
Liberati \cite{7}, and ``The Search for Quantum Gravity Signals'' by
Amelino{}-Camelia et al.\  \cite{8}. These authors conclude that at present,
there are only hints, but no compelling evidence for a Lorentz
violation from QG.}

{\selectlanguage{english}
However, Wilczek \cite{9} expressed doubts that the Planck{}-length
$L_p$ could have an important impact in many fields of
physics. This is because of the weakness of the gravitational field.
Also, Mead \cite{10} voices similar reservations.}

{\selectlanguage{english}
Observations reported in this article regarding dispersion of rather
low{}-energy gamma rays from GRBs strongly suggest that the universal
length $\ell_0$ is much larger than the
Planck{}-length $L_p$. From this, one may conclude that
physics may harbor two different minimal lengths, namely the
Planck{}-length $L_p$ and the much larger universal length
$\ell_0$.}

{\selectlanguage{english}
If  $\ell_0$ exist, it is important to have an estimate of its magnitude. 
Consequently, measurements on spectral time lags of gamma rays from GRBs may not only provide 
data on distances $L$ between GRBs origins and earth, buts also an estimate on the size of $\ell_0$.}

\section{II. Gamma ray bursts}

\indent  GRBs were first reported in the literature in 1973 \cite{11}. Since then, many
more papers on GRBs have appeared. Typically, these bursts consist of
photons with energies in a few keV to several MeV ranges. They may last
from several milliseconds to thousands of seconds. Some consist of a
few pulses, others of trains of pulses.  It is well established that GRBs
occur over cosmological distance \cite{12,13}. GRBs emit enormous amounts of
energy. Originally, it was estimated that these energies are in the
range of 10\textsuperscript{51} to 10\textsuperscript{53} erg \cite{13}.
However, this estimation might have been too high. Now, it is assumed
that these bursts originate from an accretion{}-disc, generating two
gamma{}-ray jets that propagate in opposite directions \cite{14,15}. Only if
one of these jet points to earth, gamma rays are observed. However,
shorter wavelength emission can be observed from any angle. 

{\selectlanguage{english}
No satisfactory theoretical model has been developed to explain the
enormous amount of energies emitted from these GRB sources. }

{\selectlanguage{english}
First, only gamma rays were observed. Later, afterglows were observed
that consisted of visible emission of the overheated debris, generated
by the burst. Some afterglows are observable over long time spans.}

{\selectlanguage{english}
However, halfway during the lifetime of the Compton satellite, the
Italian{}-Dutch BeppoSAX satellite discovered X{}-rays in the afterglow
that was followed by visible radiation and radiation of longer
wavelengths \cite{16}.}

{\selectlanguage{english}
Significantly, the gamma{}-rays spectrum varies during the bursts. Recording the specral lags
between channels, often, a hard (high-energy) to soft (low energy) spectral evolution is 
observed (Ford, Band and Marreson \cite{17}).
These types of lag are referred to as positive lags.}

{\selectlanguage{english}
There seems to be a consensus that gamma ray bursts occur over a very
small time span. These bursts should generate a quasi{}-thermal spectrum. However, this
type of spectrum is not observed. Because of the rather long duration
of gamma ray emission, it has been concluded 
the gamma{}-ray spectrum must be non{}-thermal. }

{\selectlanguage{english}
Accounting for the long duration of gamma ray emission, i.e., accounting
for relativistic correctness, the following assumptions is made: The
gigantic energy generated by the GRB is somehow stored within a
fireball. Relativistic synchronic shock theories have been advanced to
explain the long duration and the complex temporal spectrum of gamma
rays \cite{18,19}. Within the expanding fireball of the burst, an internal
engine produces a succession of relativistic shocks that are emitting a
stream of successive gamma{}-ray pulses, all propagating with the speed
of light $c$. Evoking some specifics of the working of the internal
engine, emission of gamma ray pulses of decreasing energy over a long
time span, gamma ray emission can be accounted for. }

{\selectlanguage{english}
However, Gonzales, Dingus, Kaneco et al.\ \cite{20} challenged the validity of
this shock theory, analyzing the gamma{}-ray spectrum of GRB941017. }

{\selectlanguage{english}
Jager, Mels, Brinkman et al.\  \cite{21} observed X{}-ray flashes from the
satellite BeppoSAX. Amati, Frontera, in `t Zand et al.\  \cite{22} reported
X{}-ray flashes from the same satellite, just before it was turned off
April 2002. That paper also contains references on earlier observations
of X{}-ray flashes. The X{}-ray flashes resemble GRBs. }

{\selectlanguage{english}
Significantly, according to observations by Lamb, Donaghy, and Graziany
\cite{23}, the satellite High Energy Transient Explorer (HETE{}-2)
provided strong evidence that the X{}-ray flashes, X{}-ray{}-rich GRBs,
and GRBs form a continuum, and therefore, that these three kinds of
bursts are the same phenomenon. This suggests that X{}-rays are also
stored in the fireball and are only emitted after gamma rays. This
raises a question. Why are X{}-rays not emitted before the onset of the
long duration of the gamma ray emission? }

{\selectlanguage{english}
In 2000, the tentative correlation on differential time lags $\Delta t$
for the arrival time of gamma rays of different energies and the
luminosity (distance) of gamma rays had been observed. According to
Norris and co{}-workers \cite{24}, gamma rays of different energies arrive
at slightly different times. The higher energy gamma rays arrive sooner
than the lower energy gamma rays. When optical red shifts were
available to determine the luminosity of GRBs, smaller time lags were
observed from gamma{}-rays originating from bursts that are closer
(high luminosity), compared to bursts with larger time lags belonging
to gamma{}-rays that originate from farther (low luminosity) origins. }

{\selectlanguage{english}
Also in 2000, Fenimore and co{}-workers \cite{25,26} observed a surprising
connection between spikiness (pulse sharpness) of gamma rays and the
luminosity of bursts. The pulse sharpness of gamma rays was more
pronounced when originating from more luminous (close) 
bursts, compared to gamma rays originating from less luminous (farther)
bursts. Here, amplitudes were smaller, but pulse widths were broader. }

{\selectlanguage{english}
Both groups suggested that these observations would allow determining
the distances L of GRBs from the observer, without performing any red
shift measurements. }

{\selectlanguage{english}
The HETE{}-2 made similar observations. This satellite detected an
exceptional bright GRB, namely GRB030329 \cite{27}. Strong emission from the
afterglow was observed, from X{}-rays to the visible spectrum. The main
features of this GRB were two strong pulses. The channels had the
energy ranges of 2{}-25 (e), 7{}-40 (d), and 7{}-80 keV (c), the pulses
were separated by about 10 s. However, the low energy X{}-ray bands at 2{}-10 keV
(f), the two pulses are separated by 7.4 s. Significantly, they are
separated by 11.2 s at the higher energy ranges of 30{}-400 keV (b).
The two pulses of different energies produced during this GRB separate
from each other during their lengthy travel through space.}

{\selectlanguage{english}
The above reported observations \cite{24,25,26,27}, namely increasing time
separation (lags) of gamma rays pulses of higher{}-energy from pulses
of lower energy implies that gamma rays of higher energy propagate
faster than their lower{}-energy counterparts. This observation
contradicts the STR.}

{\selectlanguage{english}
Presently, there is an ongoing effort in astrophysics to derive
empirical relationships describing the temporal behavior of spectral
time lags developing between gamma ray bands of different energies. }

{\selectlanguage{english}
Although the observation of positive lags seems to be the norm for long
GRBs, negative time lags have also been observed. Here, the time lag between
a leading gamma ray pulse and a second higher energy pulse is decreasing with time of travel.
This observation puts
constraints on the mechanics of how gamma ray pulses are generated. }

{\selectlanguage{english}
Ryed \cite{28} and Ryed et al.\  \cite{29} performed analytical and numerical
studies of spectral lags. Ryed \cite{28} states that the observation that
the light{}-curves in different energy bands lag behind each other is a
common feature in astronomical objects. Ryed et al.\  \cite{29} state that a
definite answer to the reason for the spectral time lags in GRBs has
not yet been given. \ \ }

{\selectlanguage{english}
Summarizing observations on GRBs, the following rough picture emerges.
One observes a string of gamma ray pulses with energies
$E_1, E_2, E_3, \ldots$, with energies $E_1>E_2>E_3>\ldots$ and with time
lags  $\Delta t_{12}, \Delta t_{23}, \Delta t_{34}, \ldots$, followed by an
X{}-ray flash. However, time lags $\Delta t_{12}, \Delta t_{23}, \Delta t_{34}$, will
increase with time (distance $L$ traveled) and energy $E$. }

\section{III. The special theory of relativity}

{\selectlanguage{english}
The special theory of relativity (STR) and quantum mechanics are the
main pillars of theoretical physics. The STR is well supported by experimental results, for example, the
Michelson{}-Morley experiment. The correct relationship between the
increase of the mass of electrons at speeds approaching the speed of
light $c$ was observed in particle accelerators, the relation $E = mc^2$.

}

The STR rests on two postulates

\liststyleWWviiiNumi
\begin{enumerate}
\item Among inertial frames, the laws of physics are equivalent. 
\item The speed of light $c$ is the same in
all frames of reference. 
\end{enumerate}

{\selectlanguage{english}
From the second postulate, the following corollary holds: Two light
signals, regardless of energies (frequencies), emitted from the same
source at different times $t_1$ and $t_2$
(time interval $\Delta t_{12}$), maintain their time
separation $\Delta t_{12}$, even when traveling over
astronomical distances.}

{\selectlanguage{english}
There are several ways to derive the STR. Here, we present the approach
used by Poincar\'{e} (1904). One postulates the invariance under coordinate
transformation of the hyperbolic wave equation:}

{\selectlanguage{english}
\begin{equation}\label{eq1} \nabla^2 \Psi - \frac{1}{c^2} \frac{\partial^2 \Psi}{\partial t^2}=0.
\end{equation}} 

{\selectlanguage{english}
Equation (\ref{eq1})  describes the propagation of electromagnetic waves,
together with most of electrodynamics. The Lorentz transformations
leave equation (\ref{eq1}) invariant. It is required that all laws of physics,
possibly with the exception of the general theory of relativity, be
invariant under Lorentz transformations.}

{\selectlanguage{english}
 The Lorentz transformations become imaginary for speeds $v>c$. From this observation, it is often argued that the
STR provides proof that superluminal velocities ($v>c$) do
not exist. However, this reasoning is incorrect. The STR implicitly
postulates that there are no superluminal velocities by requiring the
invariance of equation (1) under coordinate transformations. This is
then expressed explicitly by the Lorentz transformations becoming
imaginary for $v>c$. However, if superluminal velocities
do exist, this would only require a modification of equation (\ref{eq1}) to a
more general wave equation that also describes the superluminal
propagation of electromagnetic radiation. }

{\selectlanguage{english}
With $\Psi=\Psi_0e^{i(\omega t - kx)}$, one obtains from equation (1), $\omega^2=c^2k^2$, with $k$ the wave number. For
the group (signal) velocity: $d\omega/dk \equiv v_{gr}=c$. This leads us back to the second postulate,
namely that the group velocities $c=v_{gr}$ of all
electromagnetic signals is independent of their energies $E$.}

\bigskip

\section{IV. Extended theory}

To eliminate divergences in relativistic field equation, 
it was suggested that the STR might only be an approximate theory and an extension of the 
STR was proposed \cite{1,2}. Presently, it is assumed that possibly both, quantum mechanic as well as 
STR might be incomplete theories and it has been suggested, they should both be derivable from QG.

Length is not invariant under Lorentz
transformations. Elevating the universal length
${\ell}_0$, similar to $c$ and $h$, to a universal constant,
may lead to an ET that leads to new physics.

\begin{equation}\label{eq2} -\ell_0^2\nabla^4\Psi + \nabla^2\Psi - \frac{1}{c^2}\frac{\partial^2 \Psi}{\partial t^2}=0.\end{equation}

{\selectlanguage{english}
Equation (\ref{eq2}) extends electrodynamics. Again with $\Psi =\Psi_0 \, e^{i(\omega t - kx)}$  one obtains the dispersion relation: }

\begin{equation}\label{eq3}
\omega^2=c^2k^2(1+\ell_0^2k^2).
\end{equation}

{\selectlanguage{english}
Fujiwara \cite{30, 31, 32} proposed a somewhat more complex wave equation. Its
dispersion relation is more complex than equation (\ref{eq3}), but equation (\ref{eq3})
is obtained in first approximation from Fujiwara's dispersion
relation.}

{\selectlanguage{english}
The introduction of $\ell_0$ into this theory
introduces dispersion for electromagnetic waves. For the signal (group)
velocity of electromagnetic radiation $d\omega/dk \equiv v_{gr}$, one obtains from equation (\ref{eq3}) and $1 >> \ell_ok^2$ in approximation:}

\begin{equation}\label{eq4}
\frac{d\omega}{dk} \equiv v_{gr}=c(1+(3/2)\ell_0^2k^2).
\end{equation}

{\selectlanguage{english}
Because of $v_{gr}=c$ for the STR, one has for the time
arrival $t = L/c$ for electromagnetic signals. $L$ is the distance traveled
from the source, and the arrival time $t$ does not depend on the energy
$E(\nu)$ of the electromagnetic radiation. In ETs, however, we have $t=L/v_{gr}(E)$, and the travel time $t$ of electromagnetic
waves does depend on their energy $E(\nu)$. We have $k=2\pi / \lambda$  and $E=h\nu$. We have in good approximation $c = \lambda \nu$. We mentioned that after traveling billions of light years, some gamma rays from GRBs arrive on earth a few thousand seconds ahead of XRFs, suggesting about equal speed $c$ for both radiations. Again, using equation (\ref{eq4}): }

\begin{equation}\label{eq5}v_{gr}(E)=c(1+6\pi^2\ell_0^2E^2/c^2h^2).\end{equation}

{\selectlanguage{english}
Significantly, equation (\ref{eq2}) seems to contain new physics. Equation (\ref{eq5})
suggests that high{}-energy gamma rays will propagate faster than the
speed of ``light'' $c$. Equation (\ref{eq5}) also implies that gamma rays of
higher energy propagate faster than gamma rays of lower (soft) energy
E. With time $t$ of travel: $t=L/v_{gr}(E)$, one obtains:}

\begin{equation}\label{eq6}
t=\frac{L}{c(1+6\pi^2\ell_0^2E^2/c^2h^2)}.\end{equation}

{\selectlanguage{english}
If gamma rays of different energies $E_1$ and
$E_2$ are emitted from the same source at the time $t=0$,
they will eventually develop a time lag of $t_1-t_2=\Delta t$.

{\selectlanguage{english}
With
$6\pi^2\ell_0^2/c^2h^2=A^2$, and with $1 >>A^2E^2$
one obtains: }

\begin{equation}\label{eq7}\Delta t = A^2L(E_2^2-E_1^2)/c.\end{equation}

{\selectlanguage{english}
If $A(\ell_0)$ is known, and ${\Delta} t$ and
$E_1$ and $E_2$ are measured, $L$ can be
obtained and no red{}-shift measurements are necessary to
obtain the distance $L$ of gamma ray sources. Significantly, if $L$ is
obtained from red shift measurements, $\ell_0$ can be estimated. }

{\selectlanguage{english}
Another interesting relationship is obtained by rearranging equation (\ref{eq7}). For all radiation emitted from one GRB, the following relation
holds:}

\begin{equation}\label{eq8}\Delta t / (E_2^2-E_1^2)=A^2L/c = \text{Constant}.\end{equation}

{\selectlanguage{english}
 The derivation of above equations assumes that the dispersion relation (\ref{eq3}) holds.}

{\selectlanguage{english}
If this equation does not hold, one can try the expression:}

\begin{equation}\label{eq9}\Delta t / (E_2^m-E_1^m)=A^2L/c= \text{Constant}.\end{equation}

{\selectlanguage{english}
To measure the size of $\ell_0$ from
observations on GRBs, one would have to employ a satellite that has
many channels, each channel covering only a small energy band (range).
This is required to reduce measurement errors in $E_1$ and
$E_2$. \ These measurement errors are magnified when
forming the expressions
$E_1^2$ and $E_2^2$.}

\section{V. Physics of GRBs}
{\selectlanguage{english}
Due to the very high energies involved in the generation of GRBs,
ultra{}-relativistic conditions exist. Under these extreme conditions,
the STR may not be exactly valid, implying that the Lorentz invariance
is violated.}

{\selectlanguage{english}
We also assume that GRBs are of small duration. A cosmic event generates
a huge explosion, the GRB, which generates gamma rays at different
energies (frequencies), together with X{}-rays, UV, visible and IR, and
longer wavelength radiation. The emitted radiation may be
quasi{}-thermal, and no shock theory is required to explain the complex
temporal behaviors of radiation emitted from GRBs. }

{\selectlanguage{english}
From the relations derived on the foregoing section, the following
scenario seems to describe several observations on GRBs: }

{\selectlanguage{english}
After traveling over cosmological distances, gamma rays will experience
dispersion. The hardest (highest energy) gamma rays contained in the
burst will separate from gamma rays of lower energies. The time lag
${\Delta} t$ will increase with distance $L$ (time) of travel (positive
time lag), and line broadening might also be observed. Gamma rays of
lower energies should separate from lower energy radiation like
X{}-rays, UV, visible, IR, and longer wavelength radiation. However,
all this lower energy radiation should stay bundled together, with all
radiation traveling at the speed of light $c$. This bundled radiation
should be observable from space as an X{}-ray flash. Most likely,
X{}-ray flashes are the easiest to observe, apparently possessing the
highest energy among the bundled radiation. Every GRB produces an
X{}-ray flash. From earth, a flash of visible light might be observed
\cite{33}. Possibly, these flashes may even possess the time{}-history
(shape) of the original burst. This flash should blend into the
lingering afterglow. }

{\selectlanguage{english}
The above scenario suggests that the X{}-ray flashes could be one of the
most important objects for studying the physics of GRBs.  }

{\selectlanguage{english}
We have considerably simplified the theory of GRBs assuming
that out of nowhere, suddenly, these immense explosions occur. However,
as it has been suggested that GRBs may originate from two neutron stars
spiraling each other and getting closer and closer. During this time
period, low{}-energy gamma rays may be generated even before the main
event. Therefore, depending on the specific mechanism that generated
the GRBs, events occurring before, and even after, the main event may
also have to be considered. We define as the main event the emission
period of the most energetic gamma ray radiation. During this period,
low{}-energy gamma rays and X{}-rays are also generated.}

{\selectlanguage{english}
Besides the main scenario, there are three more sub{}-scenarios that
might be observed: }

\liststyleWWviiiNumii
\begin{enumerate}
\item From the main event, one will observe positive lags as we have discussed
in our simplified theory of GRBs.
\item Also, one would observe some pulses of higher energy gamma rays from the
main event catching{}-up with the lower{}-energy gamma rays generated
before the main event. In this case, one would observe negative
spectral lags.
\item The afterglow generated during the main event might overlap with the
low{}-energy gamma rays produced after the main event.
\item X{}-rays generated during the main event will overlap with low{}-energy
gamma rays generated after the main event. 
\end{enumerate}


\begin{thebibliography}{9}

{\selectlanguage{english}
\bibitem{1} T.G. Pavlopoulos, Phys. Rev. \textbf{159}, 1106 (1967).}

{\selectlanguage{english}
\bibitem{2} T.G. Pavlopoulos, Nuovo Cim. \textbf{60B}, 93 (1969). }

{\selectlanguage{english}
\bibitem{3} W. Heisenberg, Z. Naturforsch. \textbf{5a}, 251 (1950).}



{\selectlanguage{english}
\bibitem{4} V. A. Kostelecky and S. Samuel, Phys. Rev. D \textbf{39}, 683
(1989).}

{\selectlanguage{english}
\bibitem{5} D. Colladay and V. A. Kostelecky, Phys. Rev. \textbf{55, }6760
(1997).}

{\selectlanguage{english}
\bibitem{6} J. Magueijo, Rept. Prog. Phys. \textbf{66}, 2025 (2003).}

{\selectlanguage{english}
\bibitem{7} T. Jacobson, S. Liberaty, and D. Mattingly, hep{}-ph/0407370.}

{\selectlanguage{english}
\bibitem{8} G. Amelino{}-Camelia et al., gr{}-qc/0501053. }

{\selectlanguage{english}
\bibitem{9} F. Wilczek, Phys. Today, June 2001, p.12; November 2001, p.12;
August 2002, p.10.}

{\selectlanguage{english}
\bibitem{10} C.A. Mead, Phys. Today, November 2001, p.15.}


{\selectlanguage{english}
\bibitem{11} R. W. Klebesadel, I.B. Strong, and A. Olsen, Astrophys. J.
\textbf{182,} L85 (1973). }

{\selectlanguage{english}
\bibitem{12}  M. R. Metzger, Nature \textbf{387}, 878 (1997).}

{\selectlanguage{english}
\bibitem{13}  S. Kulkarni, et al., Proc. 5\textsuperscript{th} Huntsville GRB
Symposium 1999.}

{\selectlanguage{english}
\bibitem{14} P. Meszaros and M. J. Rees, Mon. Not. R. Astron. Soc.
\textbf{257}, 29 (1992).}

{\selectlanguage{english}
\bibitem{15} P. Merszaros and M. J. Rees, Astrophys. J. \textbf{397}, 570
(1992).}

{\selectlanguage{english}
\bibitem{16}  E. Costa et al., Nature \textbf{387}, 783 (1997). }

{\selectlanguage{english}
\bibitem{17}  L. A. Ford, D. L. Band, and J .L. Marreson, Astrophys. J.
\textbf{439}, 307 (1995).}

{\selectlanguage{english}
\bibitem{18} M. J. Rees and P. Meszaros, Mon. Not. R. Astron. Soc. \textbf{258,
4}1 (1992).}

{\selectlanguage{english}
\bibitem{19} P. Meszaros, and M. J. Rees, Astrophys. J. \textbf{405}, 278
(1993).}

{\selectlanguage{english}
\bibitem{20} M.M. Gonzalez, B.L. Dingus, Y. Kaneko et al., Nauture
\textbf{242}, 749, 2003.}

{\selectlanguage{english}
\bibitem{21} R. Jager, W.A. Mels, A.C. Brinkman, et al.,\  A\&A \textbf{125},
(1997).}

{\selectlanguage{english}
\bibitem{22} L Amati, F. Frontera, J.J.M. in `t Zand et al., A \& A \# 020427
(astro{}-ph/0407166).}

{\selectlanguage{english}
\bibitem{23} D.Q. Lamb, T.Q. Donaghy, and C. Graziani, astro{}-ph/0312634}

{\selectlanguage{english}
\bibitem{24} J. P. Norris, G. F. Marani, and J. T. Bonnell, Astrophys. J.
\textbf{534}, 248 (2000).}

{\selectlanguage{english}
\bibitem{25} B. Wu and E. E. Fenimore, Astrophys. J. \textbf{535}, L29
(2000).}

{\selectlanguage{english}
\bibitem{26} E. Ramirez Ruiz and E. E. Fenimore, Astrophys. J. \textbf{539},
712 (2000).}

{\selectlanguage{english}
\bibitem{27} R.Vanderspek et al., astro{}-ph/0401311.}

{\selectlanguage{english}
\bibitem{28} F. Ryde, A\&A in press (astro{}-ph/0411206). }

{\selectlanguage{english}
\bibitem{29} F. Ryde, D. Kocevski, \ Z. Bagoly et al., A\&A to be published
(astro{}-ph/0411219).}

{\selectlanguage{english}
\bibitem{30} K. Fujiwara, Found. Phys. \textbf{10}, 309 (1980).}

{\selectlanguage{english}
\bibitem{31} K. Fujiwara, Phys. Rev. D, \textbf{39}, 1764 (1989).}

{\selectlanguage{english}
\bibitem{32} K. Fujiwara, Gen. Relativ. Grav. (USA) \textbf{23}, 57 (1991).}

{\selectlanguage{english}
\bibitem{33} C. Akelof et al., Nature \textbf{398}, 400 (1999).}


\end{thebibliography}
\end{document}